\documentclass[structabstract]{aa}
\usepackage{graphicx}
\usepackage{txfonts}
\usepackage{epsfig}

\begin{document}
   \title{Producing type Iax supernovae from a specific class of helium-ignited WD explosions$?$}

   \author{B. Wang \inst{1,2}
          \and
          S. Justham \inst{3}
          \and
          Z. Han \inst{1,2}
          }

   \institute{Yunnan Observatories, Chinese Academy of Sciences, Kunming 650011, China;
              \email{wangbo@ynao.ac.cn, zhanwenhan@ynao.ac.cn}
              \and
              Key Laboratory for the Structure and Evolution of Celestial Objects, Chinese Academy of Sciences, Kunming 650011, China
              \and
              National Astronomical Observatories, Chinese Academy of Sciences, Beijing 100012, China;
              \email{sjustham@bao.ac.cn}
              }

   \date{Received ; accepted}

\abstract
{It has recently been proposed that one sub-class of type Ia supernovae (SNe~Ia)
is sufficiently both distinct and common to be classified separately from the bulk
of SNe Ia, with a suggested class name of ``type Iax supernovae''
(SNe~Iax), after SN 2002cx. However, their progenitors are still uncertain.}
{We study whether the population properties of this class might be understood if
the events originate from a subset of sub-Chandrasekhar mass
explosions. In this potential progenitor population, a carbon--oxygen white dwarf (CO WD) accumulates
a helium layer from a non-degenerate helium star; ignition of that
helium layer then leads to ignition of the CO WD.}
{We incorporated detailed binary evolution
calculations for the progenitor systems into a binary population synthesis
model to obtain rates and delay times for such events.}
{The predicted Galactic event rate of these explosions is $\sim$$1.5\times
10^{-3}\,{\rm yr}^{-1}$ according to our standard model, in good agreement with the measured rates of
SNe~Iax. In addition, predicted delay times are $\sim$70\,Myr$-$800\,Myr,
consistent with the fact that most of SNe Iax have been discovered
in late-type galaxies. If the explosions are assumed to be
double-detonations -- following current model expectations -- then
based on the CO WD masses at explosion
we also estimate the distribution of resulting SN brightness
($-13 \gtrsim M_{\rm bol} \gtrsim -19$~mag), which can
reproduce the empirical diversity of SNe Iax.}
{We speculate on why binaries with non-degenerate
donor stars might lead to SNe~Iax if similar systems with
degenerate donors do not. We suggest that the high mass of the helium layer
necessary for ignition at the lower accretion rates typically delivered from
non-degenerate donors might be necessary to produce SN 2002cx-like
characteristics, perhaps even by changing the nature of the CO ignition.}


\titlerunning{Type Iax supernovae from a class of He-ignited CO WD explosions$?$}

\authorrunning{B. Wang et al.}

   \maketitle

%
%________________________________________________________________

\section{Introduction} \label{1. Introduction}
Type Ia supernovae (SNe Ia) play an important role in astrophysics,
especially in the studies of cosmic evolution and galactic
chemical evolution.
Type Iax supernovae (SNe~Iax) have been proposed to form a distinct sub-class
of sub-luminous SNe Ia,  containing SNe resembling the prototype event
SN 2002cx (Li et al.\ 2003; Foley et al.\ 2013).
Those SNe~Iax are spectroscopically similar to SNe~Ia, but have lower
maximum-light velocities ($2000 \lesssim |v| \lesssim
8000$\,km\,s$^{-1}$), typically lower peak magnitudes
($-14.2 \gtrsim M_{\rm V} \gtrsim -18.9$~mag),\footnote{A typical peak
brightness of normal SNe Ia is about $-$19\,mag, and with a
spread in brightness of $\sim$1\,mag (e.g., Benetti et al.\ 2005).}
and maximum-light spectra that typically resemble those of the
bright 1991T-like events. Since the estimated rate of SNe~Iax is roughly
one third of the SN~Ia rate, they are relatively common astrophysical
events, although only 25 members of the class are currently identified
(Foley et al.\ 2013).

SNe~Iax appear not to obey the
standard luminosity-width relation of SNe~Ia (Foley et al.\
2013 and references therein); clearly this would affect any use of them as
distance indicators, although the fact that SNe~Iax are low-luminosity events
means that this is unlikely to be problematic in practice.
However, perhaps more importantly, this behaviour of SNe~Iax also
gives us an opportunity to help us understand the physics of
thermonuclear supernovae in general, since
whatever mechanism produces SNe~Iax leads to an alternative
family of lightcurve shapes from standard SNe~Ia.  Deducing the progenitors of these
explosions should help us to understand how this family of
explosions differs from standard SNe~Ia.
Observations support the supposition that SNe~Iax are from thermonuclear explosions of
carbon--oxygen white dwarfs (CO WDs), due to the evidence of C/O burning
in their maximum-light spectra (Foley et al.\ 2013).
SNe~Ia lack helium in their spectra, yet two SNe~Iax
show strong helium lines in their spectra, and so there might be helium in their
progenitor systems. However, there is no hydrogen in any SN Iax
spectra, and significantly less hydrogen than helium is typically
required to cause a signature in an SN spectra (see, e.g., Hachinger
et al.\ 2012). That spectral evidence might suggest that in the SN~Iax
progenitor systems a CO WD is accreting from a non-degenerate helium
star or a He WD.  Of those, the CO WD + He WD systems would exist in old
stellar populations as well as young populations, which is inconsistent
with the observation that most of SNe Iax have been discovered in
late-type galaxies (e.g., Valenti et al.\ 2009; Foley et al.\ 2013).\footnote{One of the
SNe Iax (i.e., SN 2008ge) was discovered in an old environment,
and the environment of the rest is very young, comparable with that of type
IIp core-collapse SNe (see Lyman et al. 2013).}
However, CO WD + He star systems are expected primarily in young
stellar populations, as has been observed. In this article, we
will investigate the population properties of CO WD + He star systems and
ask whether they are consistent with being the progenitors of SNe~Iax.

A CO WD can accrete material from a helium star to
increase its mass until it ignites near to the Chandrasekhar mass limit (e.g., Wang et al.\ 2009a).
However, standard Chandrasekhar mass explosion models have difficulty in reproducing
the low luminosities of SNe Iax (Hillebrandt \& Niemeyer 2000).
It is therefore natural to consider whether SNe Iax might be
produced by sub-Chandrasekhar mass explosions,
in which the explosion of a CO WD is triggered by the detonation of
a substantial surface layer of accreted helium (see, e.g., Nomoto 1982;
Woosley et al.\ 1986). However, the details of what happens following
burning of the helium layer are still unclear.
If helium ignites at the bottom of the helium layer, this may result in an event
known as helium double-detonation, during which one detonation wave propagates outward through
the helium layer, whereas an inward propagating pressure wave compresses
the CO core and leads to ignition followed by an outward detonation
(see, e.g., Livne 1990; Woosley \& Weaver 1994).  Helium
double-detonation sub-Chandrasekhar mass explosions have
previously been considered as promising explanations for standard SNe
Ia (see, e.g., Branch et al.\ 1995; H\"{o}flich
\& Khokhlov 1996), but modern models disagree over which known SNe, if
any, such systems might produce (see, e.g., Fink et al.\ 2007, 2010;
Kromer et al.\ 2010; Woosley \& Kasen 2011; Sim et al.\ 2012).
In principle, there is no reason why some double-detonation events
could not produce normal SNe Ia whilst other similar
progenitor systems lead to SNe Iax.

Generic double-detonation scenarios have previously been suggested for SNe~Iax,
but inconclusively (see, e.g., Foley et al.\ 2013 and references therein).
We note that Foley et al.\ (2009) suggested that an alternative
explosion model might be responsible for SNe~Iax, specifically a
failed deflagration model; this model has success in explaining the observed properties
of SNe Iax in some aspects, and could even explain the low ejecta-mass (see, e.g., Foley et al.\ 2013).
In models like this, the accreting WD would survive and potentially
possess peculiar observational properties (see also Jordan et al.\
2012; Kromer et al.\ 2013). Perhaps the main reason to doubt
double-detonation scenarios for SNe Iax is that double-detonation models
have tended not to produce the low-velocity ejecta characteristic of SNe Iax.
However, recent work has generally suggested that the diversity of
observables which result from this class of models is rather
sensitive to details of the pre-ignition conditions (see especially
Kromer et al.\ 2010; Woosley \& Kasen 2011; Sim et al.\ 2012); we
therefore see no fundamental reason why the peculiar properties of SNe Iax
might not be matched by future developments of such calculations.  In
particular, we will speculate that the relatively massive layer of
accreted helium which is expected to be present at explosion for these
progenitor systems might lead to the observed properties of SNe
Iax. Determining the nature of SN Iax progenitors should certainly help to refine
explosion models.

A second argument against double-detonations as an explanation
for SNe Iax is provided by SN 2008ha. This extreme member of the SN
Iax class was inferred to have a very low ejecta mass
($\sim$$0.3\,M_{\odot}$; Foley et al. 2010). If that ejecta mass is correct,
then this event might be explained by a He-shell
explosion (see, e.g., Bildsten et al.\ 2007; Shen \& Bildsten 2009; Waldman et al. 2011),
but a complete double-detonation seems very unlikely. In that case,
then SN 2008ha would need to be explained as a different type of
event.  However, that would not necessarily mean that this particular \emph{progenitor
population} could not have produced SN 2008ha, if ignition of the
helium shell  in these systems can lead to
less complete burning of the CO WD, perhaps even including failed
deflagrations.  Whilst recent theoretical work broadly finds that
double-detonations can be robustly triggered, several subtleties still
remain to be solved (Moll \& Woosley 2013; Shen \& Bildsten
2013). For convenience, most of this work is written as if ignition of
the helium layer in this particular progenitor population inevitably
leads to a double detonation, but the reality may well be more
complex.

There are numerous complications, including that the properties of the
helium layer that lead to helium ignition itself are expected to be a function of
CO WD mass and accretion rate (see, e.g., Bildsten et al.\ 2007; Shen
\& Bildsten 2009). Heating from differential rotation in the accreted layer
provides further uncertainty (Yoon \& Langer 2004).
Transients have also been observed which are consistent with the detonation of a thick
helium layer on a WD which does \emph{not} lead to a double-detonation
(Poznanski et al.\ 2010; Kasliwal et al.\ 2010; Perets et al. 2011). Previous studies have
often made the simplification that, for sufficiently
massive CO WDs, a helium layer with mass $0.1\,M_{\odot}$ can ignite
and lead to a double detonation (e.g., Ivanova \& Taam 2004;
Ruiter et al.\ 2011), and Fink et al.\ (2007, 2010) support the position
that layers of that mass can produce a double-detonation.

We note that SNe Iax may be a bridge between normal SNe Ia and the SN
2005E-like objects (Perets et al.\ 2010), for which helium-rich
thermonuclear explosions have also been proposed as an explanation.
These share many characteristics with SNe Iax, including luminosity,
velocity, and ejecta-mass, etc (see also Sullivan et al. 2011;
Kasliwal et al. 2012). However, these SNe appear to come from old stellar populations, in
contrast to SNe Iax.  Even more enigmatically, the known examples from
this class suggest that they are preferentially produced in the outer
regions of galaxies (see, e.g., Yuan et al. 2013; Lyman et al. 2013).

Overall, determining the progenitors of SNe~Iax could distinguish
between theoretical models which are largely separated by precise
details of thermonuclear explosion physics at WD densities.

The paper is organized as follows.
In Sect. 2, we further describe our assumptions
and binary evolution calculations. The binary evolutionary results are shown in
Sect. 3. We describe the binary population
synthesis (BPS) method in Sect. 4 and present the
BPS results in Sect. 5. Finally, a discussion and summary are given in Sect. 6.

\begin{figure*}
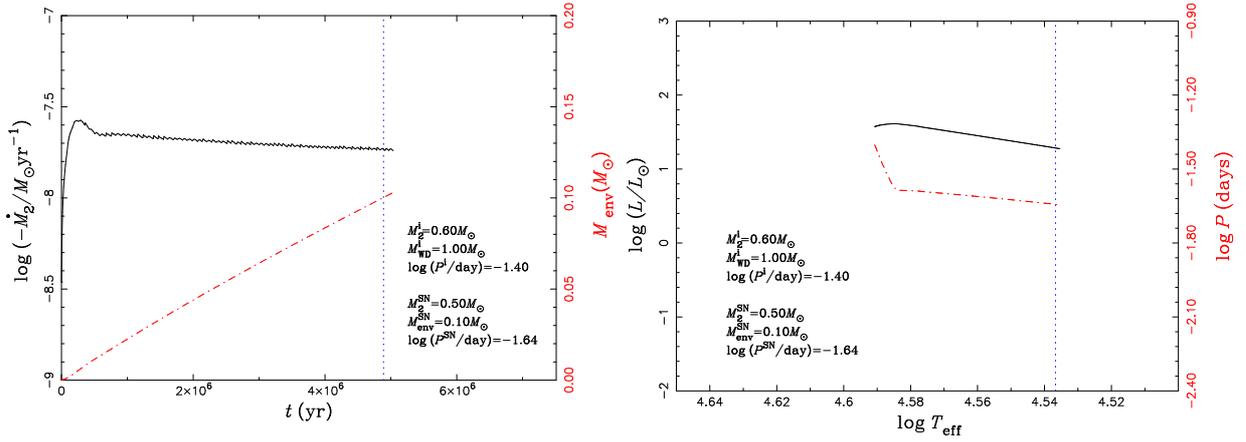

\centerline{\epsfig{file=f1a.ps,angle=270,width=8cm}\ \
\epsfig{file=f1b.ps,angle=270,width=8cm}} \caption{A representative
example of binary evolution calculations.
Left panel: the solid and dash-dotted curves show
the mass-transfer rate and the mass of the helium layer on the WD
varying with time after the helium star fills its Roche lobe, respectively.
Right panel: the evolutionary track of the
donor star is shown as a solid curve and the evolution of orbital
period is shown as a dash-dotted curve. Dotted vertical lines in
both panels indicate the position where the double-detonation may happen.
The initial binary parameters and the parameters at the moment of the SN
explosion are given in these two panels.}
\end{figure*}

\section{Binary evolution calculations}
\label{sec:binary-evolution}

Employing Eggleton's stellar evolution code (Eggleton 1971, 1972, 1973; later updated
by Han et al.\ 1994; Pols et al.\ 1998),
we have calculated the evolution of the CO WD + He star
systems. Roche lobe overflow (RLOF) is
treated within the code as described by Han et al.\ (2000).
We set the ratio of mixing length to
local pressure scale height, $\alpha=l/H_{\rm p}$, to be 2.0. In our
calculations, the initial helium star
models are composed of helium abundance $Y=0.98$ and metallicity
$Z=0.02$. Orbital angular momentum loss due to gravitational wave
radiation (GWR) is included by adopting a standard formula presented
by Landau \& Lifshitz (1971),
\begin{equation}
{d\,\ln J_{\rm GR}\over dt} = -{32G^3\over 5c^5}\,{M_{\rm WD} M_2
(M_{\rm WD}+M_2)\over a^4},
\end{equation}
where $G$, $c$, $M_{\rm WD}$ and $M_2$ are the gravitational
constant, vacuum speed of light, the mass of the accreting WD and
the mass of the companion He star, respectively.

In the double-detonation model, the helium star transfers some of its material onto the surface of
the WD, which increases the mass of the WD as a consequence.
If the mass-transfer rate onto the WD from the helium star is higher than $4\times10^{-8}\,M_{\odot}\,\mathrm{yr}^{-1}$,
stable burning allows the CO mass of the WD to increase (Woosley et al.\
1986; see also Wang et al. 2009a and references therein).  For lower accretion
rates ($1\times10^{-9}\,M_{\odot}\mathrm{yr}^{-1}\lesssim|\dot
M_2|\lesssim4\times10^{-8}\,M_{\odot}\mathrm{yr}^{-1}$),
a thick layer of helium is believed to grow on the surface of the
WD. When the mass-transfer rate drops even further ($|\dot M_2|<1\times10^{-9}\,M_{\odot} \mathrm{yr}^{-1}$),
the flash when the helium layer ignites has been suggested to be too weak
to initiate a carbon detonation, which results in only a single
helium detonation wave propagating outward (see, e.g., Nomoto et al.\
1982).\footnote{Our upper accretion rate limit only
  directly affects a small fraction of our population. The lower limit
  is uncertain but might approximately be justified in an additional way, i.e., that below
such rates the helium layer mass needed for ignition is too large to normally
be reached (see, e.g., Shen \& Bildsten 2009).}
For accretion rates of a few times
$10^{-8}\,M_{\odot}\mathrm{yr}^{-1}$, Yoon \& Langer (2004) have argued
that heating by frictional dissipation significantly reduces the
eventual chance of a helium detonation, leading to some uncertainty
in these mass-transfer rate boundaries.

According to recent hydrodynamic simulations, the minimum WD mass for
carbon burning might be $\sim$$0.8\,M_{\rm \odot}$, since the detonation of
the CO WD may be not triggered for lower mass (e.g., Sim
et al. 2012).  We also expect that the initial CO WD
masses are below $\approx$$1.1\,M_{\rm \odot}$ as more massive
WDs -- at formation -- usually consist of oxygen and neon (i.e., ONe
WDs). In principle, the WD could increase its CO mass by accreting helium as long as
the mass-transfer rate is higher than $4\times10^{-8}\,M_{\odot}\,\mathrm{yr}^{-1}$,
but this rarely occurs in our calculations (only helium donors with masses above
$\approx$$1.1\,M_{\odot}$ lead to an increase in CO mass).
Following the previous work described in the introduction, we assume that
a double-detonation occurs when a helium layer with mass
$0.1\,M_{\odot}$ accumulates on the surface of the WD.
Note, the double-detonation model seems difficult to reconcile
with the low ejecta-mass reported for SN 2008ha, for a detailed discussion see
the introduction.

We incorporated the prescriptions above into Eggleton's stellar
evolution code and followed the evolution of an ensemble of CO WD + He star
systems. The mass lost from these systems is assumed to take away the
specific orbital angular momentum of the accreting WD. We have
calculated the evolution of about 600 WD + He star systems, thereby obtaining a
large, dense model grid. The initial mass of the helium donor stars,
$M_{\rm 2}^{\rm i}$, ranges from $0.3\,M_{\odot}$ to
$1.3\,M_{\odot}$; the initial mass of the CO WDs, $M_{\rm WD}^{\rm
i}$, is from $0.8\,M_{\odot}$ to $1.10\,M_{\odot}$; the initial
orbital period of the binary systems, $P^{\rm i}$, changes from the
minimum value, at which a helium zero-age main-sequence star would fill its
Roche lobe, to $\sim$0.2\,d, where the helium star fills its Roche lobe
at the end of the helium MS.

\section{Binary evolution results} \label{3. BINARY EVOLUTION RESULTS}
\subsection{An example of binary evolution calculations}

In Fig. 1, we present an example of binary evolution calculations
for the double-detonation model.
The left panel shows the mass-transfer rate and the mass of the WD envelope varying with
time after the helium star fills its Roche lobe, whereas the right panel is the
evolutionary track of the helium donor star in the Hertzsprung-Russell
diagram, where the evolution of the orbital period is also shown.
The WD + He star binary starts with ($M_2^{\rm i}$, $M_{\rm WD}^{\rm
i}$, $\log (P^{\rm i}/{\rm day})$) $=$ (0.60, 1.0, $-$1.4), where
$M_2^{\rm i}$, $M_{\rm WD}^{\rm i}$ are the initial masses of the helium
star and of the CO WD in solar mass, and the $P^{\rm i}$ is the
initial orbital period in days.

Due to the short initial orbital period (0.04\,d),
angular momentum loss induced by GWR is large, which
leads to the rapid shrinking of the orbital separation.
After about 11\,million years, the helium star begins to fill its
Roche lobe while it is still in the helium core-burning stage.
The mass-transfer rate is stable and with a low rate of
$\sim$$2\times10^{-8}\,M_{\odot}\,\mathrm{yr}^{-1}$, which results in the formation of
a helium layer on the surface of the CO WD.  After about 5\,million
years, the mass of the helium layer increases to $0.1\,M_{\odot}$,
at which point a detonation is assumed to occur at
the base of the helium layer, and further assumed to produce a double-detonation explosion.
At this moment, the mass of the helium star is
$M^{\rm SN}_2=0.50\,M_{\odot}$ and the orbital period is $\log
(P^{\rm SN}/{\rm day})=-1.64$.

\subsection{Initial parameters for SN~Ia progenitors}

\begin{figure}
\begin{center}
\epsfig{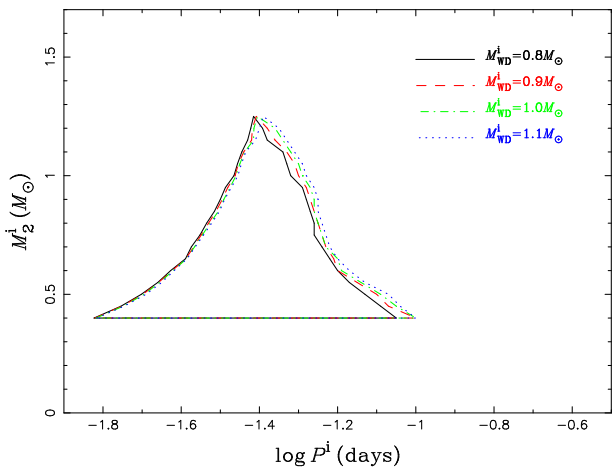} \caption{Regions in the initial
orbital period--secondary mass plane ($\log P^{\rm i}$, $M^{\rm
i}_2$) for WD + He star binaries that produce SNe Ia for various initial
WD masses. } \label{fig:contour}
\end{center}
\end{figure}

Figure 2 shows the initial contours for producing SNe Ia in
the $\log P^{\rm i}-M^{\rm i}_2$ plane for various WD masses, i.e.,
$M_{\rm WD}^{\rm i} = 0.8$, 0.9, 1.0  and $1.1\,M_{\odot}$,
where $P^{\rm i}$ and $M^{\rm i}_2$ are the initial orbital period and the initial
mass of the helium donor star, respectively.
From this figure, we can see that the contours
are slightly shifted to higher periods with the increase of initial WD masses.
This is because a helium star with a specific mass has a
smaller Roche-lobe radius with a more massive WD companion.
The left boundaries of the contours are set by the
condition that RLOF starts when the secondary is on the helium zero-age main-sequence,
whereas systems beyond the right boundary will experience a high mass-transfer rate
when the helium star evolves to the subgiant stage that is not suitable for a
double-detonation explosion. The upper boundaries are set mainly by a high
mass-transfer rate due to orbit decay induced by GWR and large
mass-ratio, which makes the WD grow in mass to the Chandrasekhar mass limit
(see Wang et al. 2009a). The lower boundaries are where the mass-transfer rate
$\dot{M}_{\rm 2}$ is higher than $1\times10^{-9}\,M_{\odot}\mathrm{yr}^{-1}$ for just long enough to produce
a $0.1\,M_{\odot}$ helium layer on the surface of the CO WD.

\section{Binary population synthesis}
\label{sec:bps}

In this double-detonation model, the progenitor systems containing a CO WD +
He star in a close binary have most likely emerged from the
common-envelope (CE) evolution of a giant binary system. CE ejection
is still an open problem.
We use the standard energy equations to calculate the output of the CE
phase (following Webbink 1984), in which there are two uncertain parameters,
$\alpha_{\rm ce}$ (the CE energetic ejection efficiency) and $\lambda$
(a structure parameter that depends on the evolutionary stage of the
donor star and the definition of the core-envelope boundary).
As in previous studies (e.g., Wang et al.\ 2010; Wang \& Han 2010), we combine $\alpha_{\rm ce}$
and $\lambda$ into a single free parameter $\alpha_{\rm ce}\lambda$,
and show results for two values: 0.5 and 1.5.

In order to obtain event rates and delay times for the double-detonation model,
we performed a series of Monte Carlo binary population
synthesis (BPS) simulations. For each BPS realization, we have used
Hurley's rapid binary evolution code (Hurley et al. 2000, 2002) to follow
the evolution of $10^{\rm 7}$ sample binaries with metallicity $Z=0.02$ from
star formation to the formation of the CO WD + He star systems based on
three evolutionary scenarios (i.e., the He star, EAGB and TPAGB
channels; for details see Wang et al.\ 2009b).  We then apply the
calculations described in Sect. 2 and assume that if the
initial parameters of a CO WD + He star system are
located inside the relevant contour of Fig. 2, a
double-detonation explosion occurs.

The Monte Carlo BPS simulations require as input the
initial mass function (IMF) of the primary, the mass-ratio
distribution, the distribution of initial orbital separations, the
eccentricity distribution of binary orbit and the star formation
rate (SFR).

(1) The IMF of Miller \& Scalo (1979, MS79) is adopted. The
primordial primary is generated according to the formula of Eggleton
et al. (1989)
\begin{equation}
M_{\rm 1}^{\rm p}=\frac{0.19X}{(1-X)^{\rm 0.75}+0.032(1-X)^{\rm
0.25}},
  \end{equation}
where $X$ is a random number uniformly distributed in the range [0,
1] and $M_{\rm 1}^{\rm p}$ is the mass of the primordial primary,
which ranges from 0.1\,$M_{\rm \odot}$ to 100\,$M_{\rm \odot}$.
The studies of the IMF by Kroupa et al. (1993) and Zoccali et al. (2000)
support this IMF.
Alternatively, we also consider the IMF of
Scalo (1986, S86)
\begin{equation} M_{\rm 1}^{\rm
p}=0.3\left(\frac{X}{1-X}\right)^{0.55},
  \end{equation}
where the meanings of $X$ and $M_{\rm 1}^{\rm p}$  are similar to
that of equation (2).

(2) The initial mass-ratio distribution of the binaries, $q'$, is
quite uncertain for binary evolution. For simplicity, we take a
constant mass-ratio distribution (Mazeh et al. 1992; Goldberg \&
Mazeh 1994),
\begin{equation}
n(q')=1, \hspace{2.cm} 0<q'\leq1,
\end{equation}
where $q'=M_{\rm 2}^{\rm p}/M_{\rm 1}^{\rm p}$.
This constant
mass-ratio distribution is supported by the study of Shatsky \&
Tokovinin (2002).
Alternatively, we also consider a rising mass
ratio distribution
\begin{equation}
n(q')=2q',\qquad  0\leq q' \leq 1,
\end{equation}
and the case in which both binary components are chosen randomly and
independently from the same IMF (uncorrelated).

(3) We assume that all stars are members of binaries and that
the distribution of separations is constant in $\log a$ for wide
binaries, where $a$ is separation and falls off smoothly at small
separation:
\begin{equation}
a\cdot n(a)=\left\{
 \begin{array}{lc}
 \alpha_{\rm sep}(a/a_{\rm 0})^{\rm m}, & a\leq a_{\rm 0},\\
\alpha_{\rm sep}, & a_{\rm 0}<a<a_{\rm 1},\\
\end{array}\right.
\end{equation}
where $\alpha_{\rm sep}\approx0.07$, $a_{\rm 0}=10\,R_{\odot}$,
$a_{\rm 1}=5.75\times 10^{\rm 6}\,R_{\odot}=0.13\,{\rm pc}$ and
$m\approx1.2$. This distribution implies that the numbers of wide
binary systems per logarithmic interval are equal, and that about
50\,percent of stellar systems have orbital periods less than
100\,yr (see, e.g., Han et al. 1995). Note, recent studies indicate that
the initial separation distribution above is reasonable for
high-mass stars (see, e.g., Sana et al. 2012), but is not good for low-mass
stars (Rahgavan et al. 2010). The SN Ia progenitor systems in this article were
surely not initially low-mass systems. Moreover, since more of the progenitor
primaries are closer to ``high-mass" than ``low-mass", then it seems more logical
to adopt the distribution for high-mass stars.

(4) A circular orbit is assumed for all binaries. The orbits of
semidetached binaries are generally circularized by the tidal force
on a timescale which is much smaller than the nuclear timescale.
Furthermore, a binary is expected to become circularized during the
RLOF. Alternatively, we also consider a uniform eccentricity
distribution in the range [0, 1].

(5) We simply assume a constant SFR over the past 14\,Gyr or,
alternatively, as a delta function, i.e., a single instantaneous starburst. In the
case of the constant SFR, we assume that one
binary with a primary more massive than $0.8\,M_{\odot}$ is formed
annually (see, e.g., Iben \& Tutukov 1984; Han et al.
1995).  For the case of the single starburst, we assume a burst producing $10^{11}\,M_{\odot}$
in stars, which is a typical mass of a galaxy. In fact, a galaxy may
have a complicated star formation history. We only choose these two
extremes for simplicity. A constant SFR is similar to the situation
of spiral galaxies (Yungelson \& Livio 1998; Han $\&$ Podsiadlowski
2004), whereas a delta function to that of elliptical galaxies or
globular clusters.

\section{Results of binary population synthesis}

\subsection{Rates and delay times of SNe Ia}

\begin{table}
 \begin{minipage}{85mm}
 \caption{Galactic SN Ia rates for different simulation sets in which set 2 is our standard model. Notes: $\alpha_{\rm ce}\lambda$ = CE ejection parameter; ${\rm IMF}$ = initial mass function;
$n(q')$ = initial mass ratio distribution; ${\rm ecc}$ = eccentricity distribution of binary
orbit; $\nu$ = Galactic SN Ia rates.}
   \begin{tabular}{cccccc}
\hline \hline
Set & $\alpha_{\rm ce}\lambda$ & ${\rm IMF}$ & $n(q')$ & ${\rm ecc}$ & $\nu$ ($10^{-3}$\,yr$^{-1}$)\\
\hline
$1$ & $0.5$     & ${\rm MS79}$   & ${\rm Constant}$  & ${\rm Circular}$ & $0.399$\\
$2$ & $1.5$     & ${\rm MS79}$   & ${\rm Constant}$  & ${\rm Circular}$ & $1.542$\\
$3$ & $1.5$     & ${\rm MS79}$   & ${\rm Constant}$  & ${\rm Uniform}$  & $1.547$\\
$4$ & $1.5$     & ${\rm S86}$    & ${\rm Constant}$  & ${\rm Circular}$ & $1.007$\\
$5$ & $1.5$     & ${\rm MS79}$   & ${\rm Rising}$    &  ${\rm Circular}$ & $1.695$\\
$6$ & $1.5$     & ${\rm MS79}$   & ${\rm Uncorrelated}$ & ${\rm Circular}$ & $0.158$\\
\hline
\end{tabular}
\end{minipage}
\end{table}

\begin{figure}
\epsfig{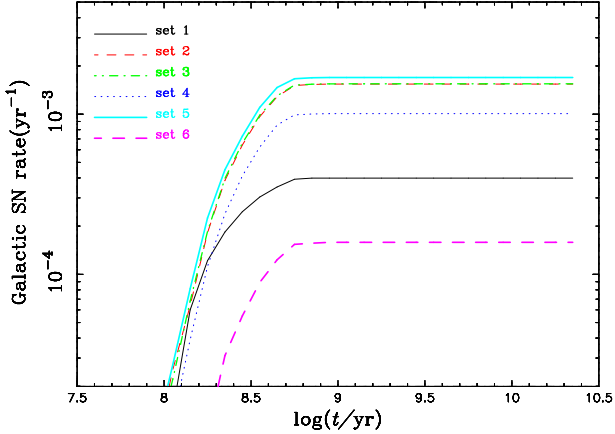} \caption{Evolution of
Galactic SN Ia rates for a constant Pop I SFR ($Z=0.02$,
${\rm SFR}=3.5\,M_{\rm \odot}{\rm yr}^{-1}$). The key to the line-styles representing different sets
is given in the upper left corner. The result of set 3
almost coincides with that of set 2.}
\end{figure}

The BPS studies are highly dependent on the chosen initial conditions in these Monte Carlo simulations.
In order to systematically investigate Galactic SN Ia rates
for the double-detonation model, six sets of simulations (see Table 1) with metallicity
$Z=0.02$ are performed, where set 2 is our standard model with
the best choice of model parameters (e.g., Wang et al. 2010b).  The
initial model parameters in other sets is varied to examine their influences on the final results.
According to the six sets of Monte Carlo simulations,
we find that the BPS is sensitive to uncertainties in some input
parameters, in particular the mass-ratio distribution. If we adopt a
mass-ratio distribution with un-correlated component masses (set 6),
the SN Ia rate will decrease significantly, as most of the donors in the double-detonation model
are not located inside the relevant contour of Fig. 2.

In Fig. 3, we show the evolution of Galactic SN Ia rates for the double-detonation model
by adopting $Z=0.02$ and ${\rm SFR}=3.5\,M_{\rm
\odot}{\rm yr}^{-1}$. The simulation for our standard model (set 2)
gives Galactic SN Ia rate of $\sim$$1.5\times 10^{-3}\,{\rm
  yr}^{-1}$ , which is roughly one third of the inferred Galactic SN Ia
rate ($3-4\times 10^{-3}\ {\rm yr}^{-1}$; Cappellaro \& Turatto 1997).
This is in good agreement with the measured rates of SNe~Iax
($31\ensuremath{^{+17}_{-13}}$ SNe~Iax for every 100 SNe~Ia in a given volume; Foley et al. 2013).
The birthrate from $\alpha_{\rm ce}\lambda=0.5$ (set 1) is lower than that of
$\alpha_{\rm ce}\lambda=1.5$ (set 2), since the post-CE binaries are more likely to
be located in the SN Ia production region for $\alpha_{\rm ce}\lambda=1.5$.
As expected, these rates are consistent with the double-detonation
model only producing part of the overall SN Ia rate (for a recent
review of other potential SN Ia formation channels see Wang \& Han 2012).

\begin{figure}[tb]
\begin{center}
\includegraphics[width=5.7cm,angle=270]{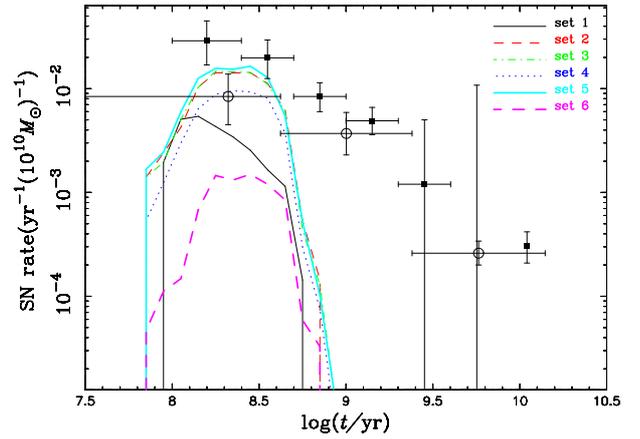}
 \caption{Delay time distributions
of SNe Ia for the double-detonation model. The filled squares and open circles are taken from Totani et al. (2008)
and Maoz et al. (2011), respectively. The result of set 3
almost coincides with that of set 2.}
  \end{center}
\end{figure}

\begin{figure}[tb]
\begin{center}
\includegraphics[width=5.7cm,angle=270]{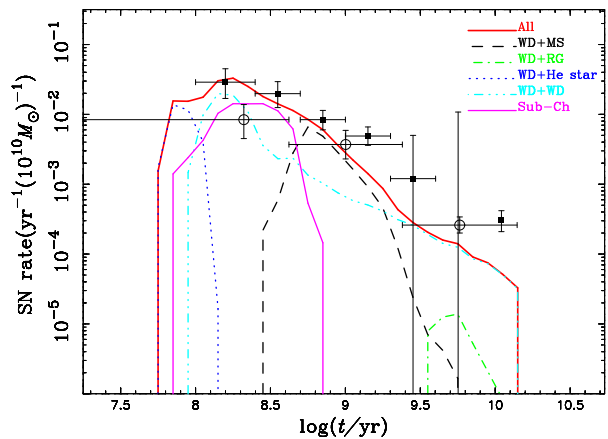}
 \caption{Delay time distributions of SNe Ia for different progenitor models.
 The results for the WD + MS, WD + RG, WD + He star and WD + WD models are
 from Wang et al. (2010b). The double-detonation sub-Chandrasekhar mass
 model presents the result of set 2.}
  \end{center}
\end{figure}

The delay times of SNe~Ia are defined as the time interval between the star formation and SN explosion.
The various progenitor models can be examined by comparing
the delay time distributions with that of observations. Fig. 4 displays
the delay time distributions of SNe Ia for the double-detonation model.
In the figure,  we see that SN Ia explosions occur between $\approx$70\,Myr
and $\approx$800\,Myr after the starburst, consistent with the current
SNe Iax sample, most of which have originated in late-type --
i.e., star-forming -- galaxies. In Fig. 5, we show the delay time distributions of SNe Ia
from different progenitor models. From this figure, we can see that the double-detonation
sub-Chandrasekhar mass model is only a subclass for producing SNe Ia.

\subsection{Initial parameters of WD + He star systems}
\vspace*{-0.5ex}
\begin{figure}[tb]
\begin{center}
\includegraphics[width=8.7cm,angle=0]{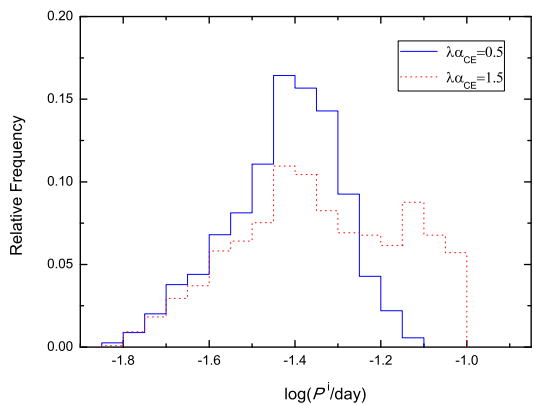}
 \caption{Distribution of the initial orbital periods of the WD + He star systems
that can ultimately produce SNe Ia. The solid and the dotted histograms
represent the cases with $\alpha_{\rm ce}\lambda=0.5$ (set 1) and
$\alpha_{\rm ce}\lambda=1.5$ (set 2), respectively. }
  \end{center}
\end{figure}

\begin{figure}[tb]
\begin{center}
\includegraphics[width=8.7cm,angle=0]{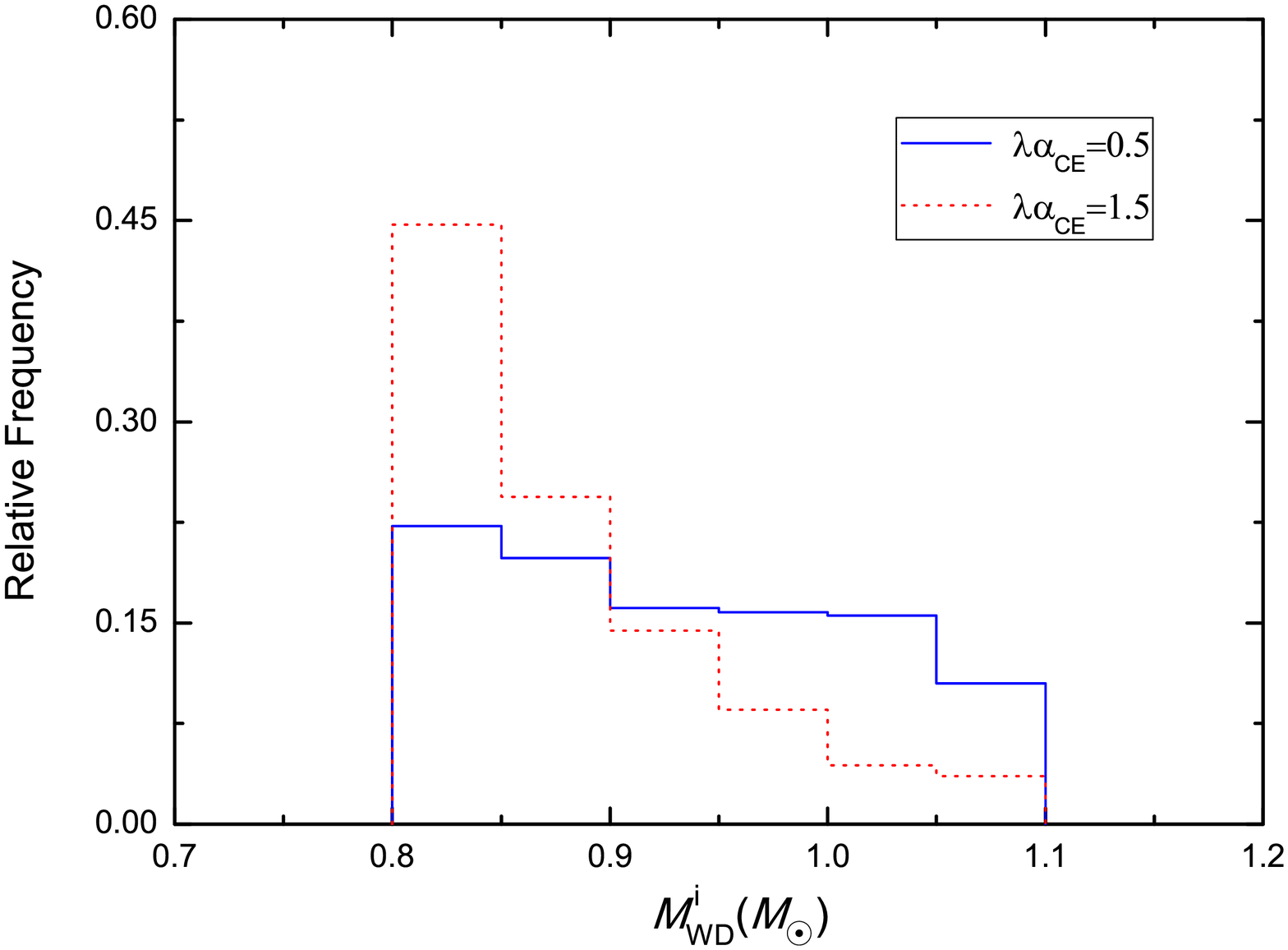}
 \caption{Similar to Fig. 6, but for the distribution of the initial WD masses in the WD + He star systems.}
  \end{center}
\end{figure}

\begin{figure}[tb]
\begin{center}
\includegraphics[width=8.7cm,angle=0]{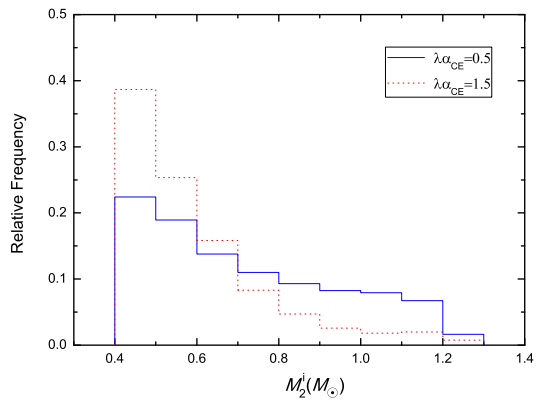}
 \caption{Similar to Fig. 6, but for the distribution of the initial masses of secondaries in the WD + He star systems.}
  \end{center}
\end{figure}

In order to aid searches for candidate SN Ia progenitors, we show some properties of
initial WD + He star systems which are predicted to produce SNe Ia based on these BPS
calculations. Fig. 6 presents the distribution of the initial orbital periods of
the WD + He star systems that ultimately produce SNe Ia with
different $\alpha_{\rm ce}\lambda$. Those distributions are given at
the current epoch, assuming a metallicity $z=0.02$ and an ongoing constant SFR.
The figure illustrates that a higher value of $\alpha_{\rm ce}\lambda$
leads to wider WD + He star binaries, as could be expected since a
high value of $\alpha_{\rm ce}\lambda$ allows the CE to be ejected
with a smaller decrease in the orbit of the binary.
Fig. 7 shows the distribution of the masses of the WDs at formation,
revealing that a low value of $\alpha_{\rm ce}\lambda$ tends to lead
to higher initial WD
masses.  This trend can be understood by considering the He star
channel (see Sect. 4), which allows stable RLOF to produce more massive WDs
(when compared to dynamical mass transfer and a CE phase). Our BPS
simulations find that a low value of $\alpha_{\rm ce}\lambda$ will
increase the fraction of SNe Ia formed via the He star channel, and
will therefore tend to produce more massive WDs.
In Fig. 8, we display the distribution of the masses of the He donor
stars as they are formed. A low value of
$\alpha_{\rm ce}\lambda$ in this figure tends to produce larger He star
masses. This is also related to the stable RLOF chanel, which
involves initially more massive companion stars and so typically leads to
larger final He-core masses (and therefore He star masses).

According to our binary evolution calculations, the majority of the binaries
has the SN ejecta-mass $\sim$($M_{\rm WD}^{\rm i}+0.1\,M_{\odot}$ helium layer mass), except
the binaries with the He donor mass $>$$1.1\,M_{\odot}$. Due to orbit decay induced
by GWR and large mass-ratio, the binaries with the
He donor mass $>$$1.1\,M_{\odot}$ experience a high mass-transfer rate, leading to
the increase of the WD mass. However, binaries with the He donor mass above
$>$$1.1\,M_{\odot}$ only account for a small proportion (see Fig. 8), i.e., they
have a small contribution for the distribution of the SN ejecta-mass.

\subsection{Luminosity distribution}

\begin{figure}[tb]
\begin{center}
\includegraphics[width=8.7cm,angle=0]{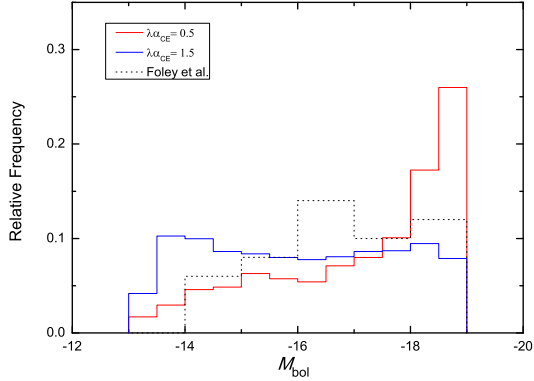}
 \caption{\label{fig:luminosity} Distribution of peak bolometric
   magnitudes from the population model (for two different $\alpha_{\rm ce}\lambda$
   values). The data are observed SN~Iax V-band magnitudes from Foley et al.\ (2013),
   several of which are lower limits.}
  \end{center}
\end{figure}

To quantify the relationship between the WD explosion mass ($M_{\rm WD}$) and the
SN Ia peak bolometric magnitude ($M_{\rm bol}$), Ruiter et al.\ (2013) recently carried out a series of simulations
for a range of WD masses based on 1D sub-Chandrasekhar mass pure
detonation models, following the earlier similar calculations by Sim et al.\ (2010).
We note that the Ruiter et al.'s numbers
are an upper limit on the luminosity, since deflagration tends to produce less
luminosity than detonation (see, e.g., Khokhlov 1991).
We also note that these simulations assumed a central detonation, with no
specified triggering mechanism; we assume that their broad results
are applicable to double-detonation events.
The SN Ia peak brightness in their simulations is directly
related to the WD explosion mass. We apply
the $M_{\rm WD}-M_{\rm bol}$ relationship
found by Ruiter et al.\ (2013) to derive the SN Ia peak brightness.

The resulting luminosity distribution is plotted in Fig.
9. There we also show observed peak V-band magnitudes from
Foley et al.\ (2013), although several of those are lower limits and
the sample is still probably too small to draw strong conclusions from
the shape of the distribution.  Nonetheless, the predicted peak
magnitude range reproduces the full
observed diversity of SNe Iax ($-14.2 \gtrsim M_{\rm V} \gtrsim
-18.9$~mag; Foley et al.\ 2013).  The predicted range ($-13 \gtrsim M_{\rm bol}
\gtrsim -19$~mag) does add a tail stretching to fainter magnitudes,
which might indicate a deficiency in the models
or an observational bias against discovering the fainter SNe Iax.
Note, these specific detonation models cannot be accurate models of SNe Iax,
and if similar detonations can somehow lead to low explosion velocities
then the physical modification which produces those low velocities would
need to leave the luminosities unchanged.

\section{Discussion and conclusions} \label{6. DISCUSSION}

\begin{figure}
\epsfig{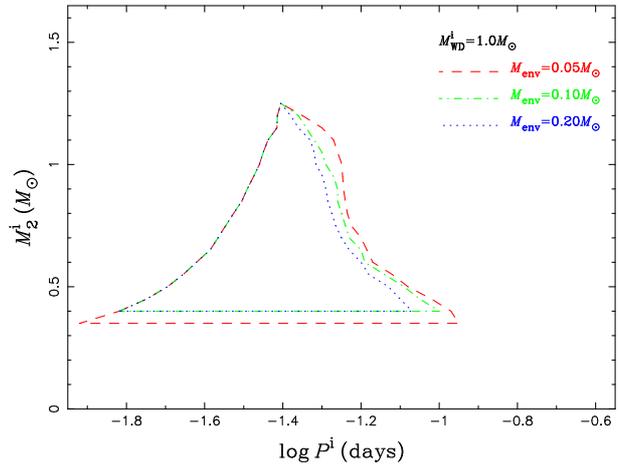} \caption{Regions in the initial
orbital period--secondary mass plane ($\log P^{\rm i}$, $M^{\rm
i}_2$) for WD + He star binaries that produce SNe Ia with initial
WD mass of $1.0\,M_{\odot}$, but for different helium layer masses. } \label{fig:contour}
\end{figure}

The regions (Fig. 2) for producing SNe Ia depend on
many uncertain input parameters, in particular the properties of the helium layer
which is poorly known. In this article, we assume that a double-detonation
is triggered when a helium layer with mass $0.1\,M_{\odot}$ accumulates on
the surface of the WD. However, it is not really known what should
be the trigger point and so, to some extent, any choice is arbitrary.
In order to check whether this choice affects the contours for producing SNe Ia,
we also show the results of 0.05 and $0.2\,M_{\odot}$ helium layer for a specific
initial WD mass (see Fig. 10). From this figure, we see that the helium
layer value has a significant influence on the contours for producing SNe Ia, and
a lower value of helium layer mass has a bigger region.
This is due to the fact that a lower value of helium layer mass needs less mass from the helium donor star.

The most relevant known system to our current work is perhaps
CD$-$30\,11223, which has been identified
as a CO WD + sdB star system with a $\sim$1.2\,h orbital period
(Vennes et al.\ 2012; Geier et al.\ 2013). Geier et al. (2013) constrained both
the mass of the sdB $\sim$$0.51\,M_{\rm\odot}$ and the mass of the WD
companion $\sim$$0.76\,M_{\rm\odot}$.
Due to the short orbital period, angular momentum loss from GWR is large.
After about 36\,million years, the sdB star will begin to fill its
Roche lobe while it is still in the He-core burning stage.
CD$-$30\,11223 might explode as an SN Ia via the double-detonation
model during its future evolution (e.g., Geier et al.\ 2013).
Other known CO WD + He-donor systems (e.g., KPD 1930+2752, V445 Pup,
and HD 49798 with its WD companion) may produce SNe Ia via other
progenitor models (see Wang \& Han 2012).

The donor star would survive the SN explosion and would potentially
be identifiable from soon after the WD is disrupted.
Identification of those remnant objects would help to support this scenario.
The surviving companions from the model might
also explain hypervelocity helium-rich stars like US 708 (e.g., Hirsch et
al.\ 2005) due to the short orbital periods at the moment of SN
explosion (see also Justham et al.\ 2009; Wang \& Han 2009).\footnote{
An alternative explanation for the origin of US 708 is the merger
model, in which US 708 was formed by the merger
of two He-WDs in a close binary which was induced as they were ejected by the interaction with the
super-massive black hole in the Galactic center (see, e.g., Hirsch et
al. 2005); the age of US 708 makes this original scenario somewhat fine-tuned. Perets (2009) recently argued
that US 708 might have been ejected as a binary from a
triple disruption by the super-massive black hole, which later on evolved and merged to
form an sdO star.}
Geier et al.\ (2013) suggested that the CO WD + sdB system CD$-$30\,11223 and the
hypervelocity star US 708 might represent two different stages of an
evolutionary sequence linked by an SN Ia explosion (the orbital velocity of the sdB at the moment of SN explosion will be
about 600\,km/s and thus close to the Galactic escape velocity).
Studying such high-velocity helium-rich stars
(and their WD descendants) might provide a way to test this model.

In this article, we have systematically studied a potential progenitor
population for SNe~Iax in which a \emph{non-degenerate} helium donor
star leads to a helium-ignited explosion of a sub-Chandrasekhar CO WD.
(1) This model can naturally produce SN explosions with helium lines and without hydrogen lines.
(2) The event rate agrees with the inferred rates of SNe~Iax.
(3) The explosions in this model occur between $\approx$70\,Myr
and $\approx$800\,Myr after the starburst, i.e., in relatively young
stellar populations, consistent with the host galaxy morphologies of most of SNe Iax.
(4) By adopting an existing relationship between CO mass and SN
luminosity for pure detonations, we find a SN luminosity range from $-$13\,mag to $-$19\,mag,
which compares well to the current diversity of SNe Iax ($-14.2 \gtrsim M_{\rm V} \gtrsim -18.9$\,mag).
The overall population properties resulting from this model therefore seem promisingly consistent
with the known collection of most of SNe~Iax.
However, the current models of double-detonation have difficulty in producing the observed
low-velocity ejecta characteristic of SNe Iax, and they cannot produce low ejecta-mass for
SN 2008ha as inferred in the introduction.

Current models typically predict that a double-detonation
explosion will follow the ignition of the He-layer, and we have
generally assumed that to be the case. However, we reiterate that there seems to be
sufficient uncertainty in those explosion models to allow other outcomes.
Our results are more suggestive than definitive,
but confirmation that helium double-detonation explosions can produce
a form of SNe~Ia would finally answer a fascinating astrophysical question,
and help to constrain our understanding of explosion physics.

If these progenitors do produce SNe Iax, it would not exclude other double-detonation
events from producing some normal SNe Ia.
However, this would suggest a new problem: why would systems with
non-degenerate helium donors make SNe~Iax but not -- or only rarely --
those with He WD donors? We can only speculate about potential answers.
Degenerate donors tend to produce higher mass-transfer
rates than our non-degenerate donors (see, e.g., Han \& Webbink 1999),
and higher mass-transfer rates require less massive helium layers for
ignition (see, e.g., Shen \& Bildsten 2009).
One possibility is that current theoretical work underestimates the
mass of helium layer which is required to lead to a
double-detonation, in which case the less massive helium-shell detonations
do not lead to a form of SNe Ia.   Alternatively -- if double-detonations are
triggered in systems with lower-mass helium layers -- then perhaps only
the more massive helium shells sufficiently alter the explosion
properties for them to be identified as peculiar, 2002cx-like,
events. Or perhaps helium layers which are \emph{too}
thick do not produce double-detonations after all; SNe Iax do require lower ejecta velocities
than the double-detonation explosion models currently predict, and perhaps this can be
achieved if massive enough helium layers can ignite CO deflagrations. We intend to study
the potential diversity of this population in future
by adopting more sophisticated assumptions about the
properties of the helium layer at ignition.
The 2002cx-like subclass of
SNe Ia certainly deserves further detailed study.

\begin{acknowledgements}
We acknowledge the anonymous referee for his/her valuable comments that helped
us to improve the paper. We also thank useful comments and suggestions from Hagai Perets.
This work is partly supported by the 973 programme of China
(No. 2014CB845703), the NSFC (Nos. 11322327, 11103072 and 11033008), the
CAS (No. KJCX2-YW-T24) and the Talent Project of Young Researchers of Yunnan province.

% (2013HB037)

\end{acknowledgements}

\clearpage

\end{document}